\documentclass[aps,preprint]{revtex4}%
\usepackage{amsfonts}
\usepackage{amsmath}
\usepackage{amssymb}
\usepackage{graphicx}%
\begin{document}
\preprint{ }
\title[Potentials over fields]{Ascendancy of potentials over fields in electrodynamics}
\author{H. R. Reiss}
\affiliation{Max Born Institute, Berlin, Germany}
\affiliation{American University, Washington, DC, USA}
\email{reiss@american.edu}


\begin{abstract}
Multiple bases are presented for the conclusion that potentials are
fundamental in electrodynamics, with electric and magnetic fields as
quantities auxiliary to the scalar and vector potentials -- opposite to the
conventional ordering. One foundation for the concept of basic potentials and
auxiliary fields consists of examples where two sets of gauge-related fields
are such that one is physical and the other is erroneous, with the information
for the proper choice supplied by the potentials. A major consequence is that
a change of gauge is not a unitary transformation in quantum mechanics; a
principle heretofore unchallenged. The primacy of potentials over fields leads
to the concept of a hierarchy of physical quantities, where potentials and
energies are primary, while fields and forces are secondary. Secondary
quantities provide less information than do primary quantities. Some criteria
by which strong laser fields are judged are based on secondary quantities,
making it possible to arrive at inappropriate conclusions. This is exemplified
by several field-related misconceptions as diverse as the behavior of charged
particles in very low frequency propagating fields, and the fundamental
problem of pair production at very high intensities. In each case, an approach
based on potentials gives appropriate results, free of ambiguities. The
examples encompass classical and quantum phenomena, in relativistic and
nonrelativistic conditions. This is a major extension of the quantum-only
Aharonov-Bohm effect, both in supporting the primacy of potentials over
fields, and also in showing how field-based conceptions can lead to errors in
basic applications.

\end{abstract}
\date[5 July 2018]{}
\maketitle

\section{Introduction}

For most of the history of exploring electromagnetic phenomena, it had been
believed that knowledge of the electric and magnetic fields in a physical
problem is sufficient to define the problem. The scalar and vector potentials,
whose spatial and temporal derivatives yield the fields, had been regarded as
auxiliary quantities that are useful but not essential. This conclusion was
apparently reinforced by the fact that the set of potentials to represent the
fields is not unique. Subject to modest restrictions, there exist
transformations (called gauge transformations) to other sets of potentials
that produce the same fields.

This seemingly straightforward situation was upset by the Aharonov-Bohm effect
\cite{es,ab}. The simplest realization of this phenomenon is that an electron
beam passing outside a solenoid containing a magnetic field will be deflected,
even though there is no field outside the solenoid. There is, however, a
potential outside the solenoid that suffices to explain the deflection. The
effect remained controversial until it was verified experimentally
\cite{tonomura}. This has the basic consequence that potentials are more
fundamental than fields. The Aharonov-Bohm effect is founded on a single
explicitly quantum-mechanical phenomenon, and commentary about its
significance has been in terms of quantum mechanics \cite{furry,vaidman}.

The concept explored here is different from that of the Aharonov-Bohm effect,
and much more consequential. It is shown that a change of gauge can introduce
a violation of basic symmetries, even when the usual constraints on allowable
gauge transformations have been satisfied. Furthermore, these symmetry
violations can occur in classical physics as well as in quantum mechanics with
external electromagnetic fields. The consequences of these results are
profound. There exist contrasting sets of potentials that yield exactly the
same fields, but where one set is consonant with physical requirements but
another is not. This proves directly that the selection of the proper set of
potentials is the decisive matter, since the predicted fields are the same in
both cases. A corollary is that gauge transformations are not unitary
transformations. This contradicts the field-based assumption that gauge
transformations must preserve the values of measurable quantities. The
assumption of unitarity (often implicit) underlies some of the influential
articles that have been published on the subject of gauge choice.

Examples employed here to demonstrate the primacy of potentials -- a charged
particle in interaction with a constant electric field, and a bound electron
subjected to a plane-wave field -- represent basic physics problems, unlike
the narrow specificity of the Aharonov-Bohm effect. An important feature
revealed by these examples is that, although the physical consequences of
static or quasistatic-electric (QSE) fields are quite similar to plane-wave
effects at the low field intensities that exist in the usual atomic, molecular
and optical (AMO) physics processes, at the high field intensities now
achievable with laser fields, they can be profoundly different. These
differences have yet to be fully appreciated in the AMO literature, leading to
misconceptions that persist nearly forty years after the first laboratory
observation \cite{ati} of explicit intense-field effects.

A concept introduced here is that of a hierarchy of physical quantities. Since
potentials are primary and fields are secondary, it follows that energies are
primary and forces are secondary. This ranking resolves the long-standing
mystery about why the Schr\"{o}dinger equation cannot be written directly in
terms of electric and magnetic fields, even though the fields were
conventionally assumed to be basic physical quantities. All attempts to
express the Schr\"{o}dinger equation directly in terms of fields have resulted
in nonlocality \cite{mandelstam,dewitt,belinfante,levy,rohrlich,priou}. This
apparent anomaly is one of the enduring puzzles of quantum mechanics. A
hierarchy of physical quantities also serves to clarify the current confused
situation in the strong-laser community, where field-based intensity measures
are employed that are inconsistent with energy-based criteria. One important
example of this misdirection is the introduction of the concept of the
\textquotedblleft critical electric field\textquotedblright%
\ \cite{sauter,schwinger} into the discussion of strong laser effects, despite
the fact that lasers produce transverse fields and the critical field has
well-defined meaning only for longitudinal fields. The basic differences
between transverse fields and longitudinal fields are also exhibited in the
macroscopic world in terms of the properties of extremely-low-frequency radio waves.

The range of applicability of the concepts examined here is very large, since
it encompasses classical electromagnetism, and also relativistic and
nonrelativistic quantum mechanics in which the electromagnetic field is
regarded as an external classical field.

The limitation to external classical fields is significant, since it places
the present work outside the scope of a Yang-Mills theory \cite{yangmills}.
Quantum electrodynamics (QED) is a Yang-Mills theory, but standard QED does
not incorporate strong-field theory. Strong-field theories contain an apparent
intensity-dependent \textquotedblleft mass shift\textquotedblright, discovered
independently by Sengupta \cite{sengupta} and by the present author
\cite{hrdiss,hr62}. This mass shift can be explained in terms of a demand for
covariance in external fields \cite{hrup}. The mass shift is a fundamental
phenomenon in strong-field physics \cite{hrje}, but it does not exist in the
context of the quantized fields of QED \cite{hrdiss,hr62,jehr}.

Section II below discusses two basic examples that exhibit pairs of potential
choices that describe exactly the same fields, but where one set of potentials
is physically acceptable and the other is not. One example is the simplest
possible case: the classical interaction of a charged particle with a constant
electric field. Of the two possibilities for gauge choice, one contradicts
Noether's Theorem \cite{noether}. There is no such problem with the
alternative gauge. The next example is the interaction of a charged particle
with a plane-wave field, such as the field of a laser. In this case, the key
factor is that the symmetry principle in question -- preservation of the
propagation property of a plane-wave field -- is not often mentioned, even in
the context of very strong fields where this symmetry is crucial \cite{hrup}.
The demand for the preservation of the propagation property imposes a strong
limitation on possible gauge transformations. In the presence of a
simultaneous scalar interaction, like a Coulomb binding potential, only the
radiation gauge is possible \cite{hrgge}. This limitation to a unique gauge
exists in both classical and quantum domains. An important aspect of this
problem is that the widely-used dipole approximation in the description of
laser-caused effects suppresses this symmetry, thus masking the errors that
follow from ignoring this basic property. Both the constant-electric-field and
the propagating-field examples admit of only one possible gauge. This lack of
gauge-equivalent alternatives is extremely important, since both situations
represent commonplace physical environments. This is in contrast to the
specialized Aharonov-Bohm effect.

An immediate consequence of the demonstrated fact that some nominally valid
gauge transformations can have unphysical consequences is that a gauge
transformation is not a unitary transformation. This is discussed in Section
III, where it is shown to be related to the construction of exact transition amplitudes.

The impossibility of writing the Schr\"{o}dinger equation directly in terms of
electric and magnetic fields is discussed in Section IV. This is further
evidence of the basic nature of potentials, and it also supports the notion of
a hierarchy of physical phenomena. Quantum mechanics can be constructed from
classical mechanics when expressed in terms of system functions like the
Lagrangian or Hamiltonian, whereas a Newtonian form of classical mechanics has
such an extension only by extrapolation to a desired result. System functions
are related to energies, whereas Newtonian physics involves forces, and forces
are directly connected with electric and magnetic fields, as shown by the
Lorentz force expression.

The ambiguities inherent in the view that the $\mathbf{E}^{2}-\mathbf{B}^{2}$
and $\mathbf{E}\cdot\mathbf{B}$ Lorentz invariants reliably characterize the
electrodynamic environment is another topic examined in Section IV. (The
Lorentz invariants, as are all electromagnetic quantities throughout this
paper, are stated in Gaussian units.) This concept has an important failure
when both invariants are zero, since it associates propagating plane-wave
fields with the completely different constant crossed fields. The
commonly-held assumption that constant crossed fields are a zero-frequency
limit of plane-wave fields (see, for example, Refs. \cite{nikrit66,ritus85}),
is shown to be untenable.

Another topic in Section IV is the apparent dominance of the electric
component of the Lorentz force expression at low frequencies, a field-related
conception that draws attention away from the rising importance of the
magnetic component of a propagating field as the frequency declines
\cite{hr75,hr82}. Inappropriate emphasis on the electric field has caused
conceptual errors even in relativistic phenomena, as discussed in Section IV
in the context of vacuum pair production. A potentials-related approach
obviates this electric-field-dominance hazard. The concept of the
\textit{critical field} is often mentioned in connection with strong-laser
interactions \cite{arb,ssb}. The critical field refers to that value of
electric field at which spontaneous pair production from the vacuum becomes
significant. It has been applied to laser fields in terms of the electric
component of a plane-wave field. This is devoid of meaning for laser beams in
vacuum because pure electric fields and plane-wave fields are disjoint
concepts, as is evident from Section II. The conservation conditions
applicable to critical-field considerations cannot be satisfied by a laser.
Even were the electric component of a laser field equal to the critical field,
pair production cannot occur because pair production from the vacuum by a
laser pulse cannot occur unless there is a counter-propagating field to
provide the necessary conservation of momentum \cite{hrdiss,hr62,hr71,burke}.
The fact that photons convey momentum is incompatible with the concept of a
critical electric field for laser-induced processes.

Section IV concludes with the practical problem of communicating with
submerged submarines. This has been done under circumstances that emphasize
how different plane-wave fields are from QSE fields

Section V explores the notion of a hierarchy of physical quantities.
Potentials are directly related to energies, so they are identified as primary
quantities. Fields are derived from potentials, so they are secondary. Forces
are determined by fields and so forces are also secondary. The hierarchy
concept is related to classical mechanics in that Newtonian physics is couched
in terms of forces, and so it is secondary to versions of classical mechanics
based on energy-based system functions like the Lagrangian and Hamiltonian.
Mechanics formulated with system functions infer Newtonian mechanics, but the
converse is not true.

\section{Symmetry violation}

The two examples presented have an important qualitative difference. The first
example -- a constant electric field -- is so elementary that the proper
choice of potentials is obvious, and there is no motivation to explore the
properties of the symmetry-violating alternative potentials. The next example
is quite different in that the improper choice of potentials is very
attractive to a laser-physics community that is accustomed to the dipole
approximation. The requisite propagation property never appears within the
dipole approximation, and its violation is thereby invisible.

A preliminary step is to introduce the units and conventions employed in this
article, and to add some general remarks about terminology.

\subsection{Units and conventions}

Gaussian units are employed for all electromagnetic quantities. The
expressions for the electric field $\mathbf{E}$ and magnetic field
$\mathbf{B}$ in terms of the scalar potential $\phi$ and the 3-vector
potential $\mathbf{A}$ are%
\begin{equation}
\mathbf{E}=-\mathbf{\nabla}\phi\mathbf{-}\frac{1}{c}\partial_{t}%
\mathbf{A},\qquad\mathbf{B}=\mathbf{\nabla\times A.} \label{a}%
\end{equation}
A gauge transformation generated by the scalar function $\Lambda$ is%
\begin{equation}
\widetilde{\phi}=\phi+\frac{1}{c}\partial_{t}\Lambda,\qquad\widetilde
{\mathbf{A}}=\mathbf{A-\nabla}\Lambda, \label{b}%
\end{equation}
where $\Lambda$ must satisfy the homogeneous wave equation%
\begin{equation}
\left(  \frac{1}{c^{2}}\partial_{t}^{2}-\mathbf{\nabla}^{2}\right)
\Lambda=\partial^{\mu}\partial_{\mu}\Lambda=0. \label{c}%
\end{equation}
Relativistic quantities are expressed with the time-favoring Minkowski metric,
with the signature $\left(  +---\right)  $, where the scalar product of two
4-vectors $a^{\mu}$ and $b^{\mu}$ is%
\begin{equation}
a\cdot b=a^{\mu}b_{\mu}=a^{0}b^{0}-\mathbf{a\cdot b}. \label{e}%
\end{equation}
The 4-vector potential $A^{\mu}$ incorporates the scalar and 3-vector
potentials as%
\begin{equation}
A^{\mu}:\left(  \phi,\mathbf{A}\right)  . \label{g}%
\end{equation}
In 4-vector notation, the two gauge transformation expressions in Eq.
(\ref{b}) become the single expression%
\begin{equation}
\widetilde{A}^{\mu}=A^{\mu}+\partial^{\mu}\Lambda. \label{h}%
\end{equation}
Both the initial and gauge-transformed 4-vector potentials must satisfy the
Lorenz condition%
\begin{equation}
\partial^{\mu}A_{\mu}=0,\qquad\partial^{\mu}\widetilde{A}_{\mu}=0. \label{j}%
\end{equation}
The propagation 4-vector $k^{\mu}$ consists of the propagation 3-vector
$\mathbf{k}$ as the space part, and the amplitude $\left\vert \mathbf{k}%
\right\vert =\omega/c$ as the time component:%
\begin{equation}
k^{\mu}:\left(  \omega/c,\mathbf{k}\right)  . \label{k}%
\end{equation}
The 4-vector $k^{\mu}$ defines the light cone and, according to the rule
(\ref{e}), it is \textquotedblleft self-orthogonal\textquotedblright:%
\begin{equation}
k^{\mu}k_{\mu}=\left(  \omega/c\right)  ^{2}-\mathbf{k}^{2}=0, \label{l}%
\end{equation}
which is an important possibility in this non-Euclidean space.

The concept of transversality refers to the property of plane-wave fields
expressed in a relativistic context as \textit{covariant transversality}%
\begin{equation}
k^{\mu}A_{\mu}=0, \label{m}%
\end{equation}
in terms of the 4-potential $A^{\mu}$. In many textbooks on classical
electromagnetic phenomena, transversality is defined as \textit{geometrical
transversality}%
\begin{equation}
\mathbf{k\cdot E}=0\text{ and }\mathbf{k\cdot B}=0, \label{n}%
\end{equation}
in terms of the electric and magnetic fields. It can be shown that covariant
transversality infers geometrical transversality.

\subsection{Terminology}

Despite the conclusion in this paper that potentials are more basic than
fields, it is not possible to avoid the use of the term \textquotedblleft
field\textquotedblright\ in a generic sense. For example, one important
conclusion reached herein is that vector and scalar potentials provide more
information than do electric and magnetic fields in the description of the
effects of laser fields. In the preceding sentence, the term \textquotedblleft
laser field\textquotedblright\ is used generically to identify the radiation
created by a laser, despite the particular result that potentials are the
better approach in the description of that radiation. A similar problem arises
when it is concluded that the dipole approximation amounts to the replacement
of the \textquotedblleft transverse field\textquotedblright\ of a laser by the
more elementary \textquotedblleft longitudinal field\textquotedblright. In
each of the phrases demarcated by quotation marks, the word \textquotedblleft
field\textquotedblright\ is used in a generic sense to identify an
electromagnetic phenomenon.

\subsection{Constant electric field}

The problem of a particle of mass $m$ and charge $q$ immersed in a constant
electric field of magnitude $E_{0}$ is inherently one-dimensional. For present
purposes, nothing is gained by going to three spatial dimensions. The problem
is clearly one in which energy is conserved. By Noether's Theorem
\cite{noether}, the Lagrangian must be independent of time $t$, so that the
connection between the electric field and potentials given in Eq. (\ref{a})
must depend only on the scalar potential $\phi.$ Equation (\ref{a}) can then
be integrated to give the potentials%
\begin{equation}
\phi=-xE_{0},\qquad A=0, \label{o}%
\end{equation}
since an additive constant of integration has no physical meaning. The
potentials descriptive of this problem are unique, and given by Eq. (\ref{o}).

The Lagrangian function is the difference of the kinetic energy $T$ and the
potential energy $U$:%
\begin{align}
L  &  =T-U\label{p}\\
&  =\frac{1}{2}m\overset{.}{x}^{2}+qxE_{0}. \label{p1}%
\end{align}
The Lagrangian equation of motion is%
\begin{equation}
\frac{d}{dt}\frac{\partial L}{\partial\overset{.}{x}}-\frac{\partial
L}{\partial x}=m\overset{..}{x}-qE_{0}=0, \label{q}%
\end{equation}
which is just the elementary Newtonian equation%
\begin{equation}
m\overset{..}{x}=qE_{0}. \label{r}%
\end{equation}
The simplest initial conditions for this problem -- initial position and
velocity set to zero -- lead to the solution%
\begin{equation}
x=\frac{qE_{0}}{2m}t^{2}. \label{s}%
\end{equation}
From Eqs. (\ref{p}) and (\ref{p1}), it follows that%
\begin{equation}
T=\frac{1}{2m}\left(  qE_{0}t\right)  ^{2},\qquad U=-\frac{1}{2m}\left(
qE_{0}t\right)  ^{2},\qquad T+U=0. \label{t}%
\end{equation}
The anticipated conservation of energy holds true.

Despite the uniqueness of the potentials of Eq. (\ref{o}), there exists an
apparently proper gauge transformation generated by the function%
\begin{equation}
\Lambda=ctxE_{0}. \label{u}%
\end{equation}
The gauge-transformed potentials are%
\begin{equation}
\widetilde{\phi}=0,\qquad\widetilde{A}=-ctE_{0}, \label{v}%
\end{equation}
and the Lagrangian function is \cite{hrjmo}%
\begin{equation}
\widetilde{L}=\frac{1}{2}m\overset{.}{x}^{2}-qtE_{0}\overset{.}{x}. \label{w}%
\end{equation}
The kinetic energy is unaltered ($\widetilde{T}=T$), but the new potential
energy is%
\begin{equation}
\widetilde{U}=qtE_{0}\overset{.}{x}, \label{x}%
\end{equation}
which is explicitly time-dependent. The new equation of motion is%
\begin{equation}
\frac{d}{dt}\left(  \frac{\partial\widetilde{L}}{\partial\overset{.}{x}%
}\right)  -\frac{\partial\widetilde{L}}{\partial x}=m\overset{..}{x}-qE_{0}=0,
\label{y}%
\end{equation}
which is identical to that found in the original gauge, so that the solution
is the same as Eq. (\ref{s}). However, the altered gauge has introduced a
fundamental change. The gauge-transformed potential energy is evaluated as%
\begin{equation}
\widetilde{U}=\frac{1}{m}\left(  qE_{0}t\right)  ^{2}, \label{z}%
\end{equation}
so that
\begin{equation}
\widetilde{T}+\widetilde{U}=\frac{3}{2m}\left(  qE_{0}t\right)  ^{2}.
\label{aa}%
\end{equation}
The total energy is not conserved, as was presaged by the explicit time
dependence of the gauge-transformed Lagrangian (\ref{w}).

How did this happen? One constraint placed on gauge transformations (see, for
example, the classic text by Jackson \cite{jackson}) is that the generating
function must be a scalar function that satisfies the homogeneous wave
equation, as in Eq. (\ref{c}). This is satisfied by the function (\ref{u}).
The only other condition is the Lorenz condition (\ref{j}), which is satisfied
by the potentials before and after transformation. However, there is no
condition that guarantees preservation of symmetries inherent in the physical
problem. It is not enough to employ appropriate fields; it is necessary to
employ the appropriate potentials to ensure that all aspects of the physical
problem are rendered properly.

This writer is unaware of any instance where inappropriate potentials have
been accepted and employed in this exceedingly simple problem. The same cannot
be said for the next example.

\subsection{Plane-wave field}

Laser fields are of central importance in contemporary physics, and laser
fields are plane-wave fields. A plane-wave field is the only electromagnetic
phenomenon that has the ability to propagate indefinitely in vacuum without
the continued presence of sources. In the typical laboratory experiments with
lasers, the practical consequence of this ability to propagate without need
for sources means that all fields that arrive at a target can only be a
superposition of plane-wave fields. Any contamination introduced by optical
elements like mirrors or gratings can persist for only a few wavelengths away
from such elements. On the scale of a typical laboratory optical table, this
is negligible.

Plane-wave fields propagate at the speed of light in vacuum; they are
fundamentally relativistic. The 1905 principle of Einstein is basic: the speed
of light is the same in all inertial frames of reference \cite{einstein}. The
mathematical statement of this principle is that any description of a
plane-wave field can depend on the spacetime coordinate $x^{\mu}$ only as a
scalar product with the propagation 4-vector $k^{\mu}$. The consequence of
this projection of the spacetime 4-vector onto the light cone is that any
change of gauge must be such as to be confined to the light cone. That is,
with the definition%
\begin{equation}
\varphi\equiv k^{\mu}x_{\mu}, \label{ab}%
\end{equation}
the field 4-vector must be such that
\begin{equation}
A_{pw}^{\mu}=A_{pw}^{\mu}\left(  \varphi\right)  , \label{ac}%
\end{equation}
where the subscript \textit{pw} stand for \textit{plane-wave}. When the gauge
transformation of Eq. (\ref{h}) is applied, the gauge-altered 4-vector
potential is confined by the condition (\ref{ac}) to the form \cite{hrgge}%
\begin{equation}
\widetilde{A}^{\mu}=A^{\mu}+k^{\mu}\Lambda^{\prime}, \label{ad}%
\end{equation}
where the gauge-change generating function can itself depend on $x^{\mu}$ only
in the form of $\varphi,$ and
\begin{equation}
\Lambda^{\prime}=\frac{d}{d\varphi}\Lambda\left(  \varphi\right)  . \label{ae}%
\end{equation}
As is evident from Eq. (\ref{l}), transversality is maintained by the gauge
transformation (\ref{ad}).

A further limitation arises if an electron is subjected to a scalar binding
potential in addition to the vector potential associated with the laser field.
A relativistic Hamiltonian function for a charged particle in a plane-wave
field contains a term of the form%
\begin{equation}
\left(  i\hslash\partial^{\mu}-\frac{q}{c}A^{\mu}\right)  \left(
i\hslash\partial_{\mu}-\frac{q}{c}A_{\mu}\right)  . \label{ae1}%
\end{equation}
This occurs in the classical case, in the Klein-Gordon equation of quantum
mechanics, and in the second-order Dirac equation of quantum mechanics
\cite{feynmangm,schweber}. The expansion of the expression in Eq. (\ref{ae1})
contains the squared time part%
\begin{equation}
\left(  i\hslash\partial_{t}-\frac{q}{c}A^{0}\right)  ^{2}. \label{ae2}%
\end{equation}
If $A^{0}$ contains contributions from both a scalar potential and the time
part of the plane-wave 4-vector potential, then executing the square in Eq.
(\ref{ae2}) would give a term containing the product of these two scalar
potentials that is not physical; it does not occur in the reduction of
relativistic equations of motion to their nonrelativistic counterparts
\cite{hrgge}. This applies specifically to applications in AMO physics. That
is, it must be true that \cite{hr79,hrgge}%
\begin{equation}
A_{pw}^{0}=\phi_{pw}=0. \label{af}%
\end{equation}
This means that gauge freedom vanishes. Only the \textit{radiation gauge}
(also known as \textit{Coulomb gauge}) is possible. This is the gauge in which
scalar binding influences are described by scalar potentials $\phi$ and laser
fields are described by 3-vector potentials $\mathbf{A}$.

Consider the gauge transformation generated by the function \cite{hr79}%
\begin{equation}
\Lambda=-A^{\mu}\left(  \varphi\right)  x_{\mu}. \label{ag}%
\end{equation}
This leads to the transformed gauge%
\begin{equation}
\widetilde{A}^{\mu}=-k^{\mu}x^{\nu}\left(  \frac{d}{d\varphi}A_{\nu}\right)  ,
\label{ah}%
\end{equation}
which was introduced in Ref. \cite{hr79} in an attempt to base the Keldysh
approximation \cite{keldysh} on plane-wave fields rather than on quasistatic
electric fields. The transformed 4-potential can also be written as
\cite{hr79}%
\begin{equation}
\widetilde{A}^{\mu}=-\frac{k^{\mu}}{\omega/c}\mathbf{r\cdot E}\left(
\varphi\right)  , \label{ai}%
\end{equation}
thus suggesting a relativistic generalization of the nonrelativistic
\textit{length gauge} used by Keldysh and widely employed within the AMO
community. The problem with the $\widetilde{A}^{\mu}$ of Eq. (\ref{ah}) or
(\ref{ai}) is that it violates the symmetry (\ref{ac}) required of a
propagating field. Nevertheless, this $\widetilde{A}^{\mu}$ satisfies the
Lorenz condition (\ref{j}) and the transversality condition (\ref{m}); and the
generating function of Eq. (\ref{ag}) satisfies the homogeneous wave equation
of Eq. (\ref{c}) \cite{hr79}. That is, all the usual requirements for a gauge
transformation are met even though the transformed 4-vector potential
$\widetilde{A}^{\mu}$ of Eq. (\ref{ah}) or (\ref{ai}) violates the symmetry
required of a propagating field like a laser field.

This violation of a basic requirement for a laser field has unphysical and
hence unacceptable consequences. The most obvious is that the covariant
statement of the all-important \cite{hrup} ponderomotive energy $U_{p}$
produces a null result since%
\begin{equation}
\widetilde{U}_{p}\sim\widetilde{A}^{\mu}\widetilde{A}_{\mu}=0 \label{aj}%
\end{equation}
as a consequence of the self-orthogonality of the propagation 4-vector
$k^{\mu}$. The resemblance of Eq. (\ref{ai}) to the length-gauge
representation of a quasistatic electric field suggests a tunneling model for
the relativistic case \cite{vspopov,heidelberg}, which is inappropriate for
strong laser fields. Tunneling can occur only through interference between
scalar potentials, and a strong laser field is inherently vector, not scalar.

The basic defect of the potentials (\ref{ah}) or (\ref{ai}) is violation of
the Einstein condition of the constancy of the speed of light in all Lorentz
frames, despite the validity of the gauge transformation leading to those
potentials. The importance of the physical situation in which this occurs is
robust evidence of the significance of the proper choice of potentials, since
the electric and magnetic fields attained from the unacceptable potentials
(\ref{ah}) or (\ref{ai}) are exactly the same as those that follow from
potentials that satisfy properly the condition (\ref{ac}).

\section{Gauge transformations and unitarity}

Unitary transformations in quantum physics preserve the values of physical
observables. It was shown above that not all gauge transformations produce
physically acceptable results. Therefore, gauge transformations are not
unitary transformations. This conclusion is supported by the basic structure
of transition amplitudes.

Transition amplitudes without resort to perturbation theory are best expressed
by S matrices. These are of two (equivalent) types. The \textit{direct-time}
or \textit{post }amplitude is%
\begin{equation}
\left(  S-1\right)  _{fi}=-\frac{i}{\hslash}\int_{-\infty}^{\infty}dt\left(
\Phi_{f},H_{I}\Psi_{i}\right)  , \label{aq}%
\end{equation}
and the \textit{time-reversed} or \textit{prior} amplitude is%
\begin{equation}
\left(  S-1\right)  _{fi}=-\frac{i}{\hslash}\int_{-\infty}^{\infty}dt\left(
\Psi_{f},H_{I}\Phi_{i}\right)  . \label{ar}%
\end{equation}
The indices $f$ and $i$ label the final and initial states. The $\Phi$ states
are non-interacting states and the $\Psi$ states are fully interacting states
satisfying, respectively, the Schr\"{o}dinger equations%
\begin{align}
i\hslash\partial_{t}\Phi &  =H_{0}\Phi,\label{as}\\
i\hslash\partial_{t}\Psi &  =\left(  H_{0}+H_{I}\right)  \Psi, \label{at}%
\end{align}
where $H_{I}$ is the interaction Hamiltonian.

In a gauge transformation, the matrix elements within the time integrations in
Eqs. (\ref{aq}) and (\ref{ar}) transform as%
\begin{align}
\left(  \Phi_{f},H_{I}\Psi_{i}\right)   &  \rightarrow\left(  \Phi
_{f},\widetilde{H}_{I}\widetilde{\Psi}_{i}\right)  ,\label{au}\\
\left(  \Psi_{f},H_{I}\Phi_{i}\right)   &  \rightarrow\left(  \widetilde{\Psi
}_{f},\widetilde{H}_{I}\Phi_{i}\right)  . \label{av}%
\end{align}
Because the noninteracting states are unaltered in a gauge transformation,
there is no necessary\textit{ }equivalence between the two sides of the
expressions in Eqs. (\ref{au}) and (\ref{av}).

Those authors that endorse the favored status of the length gauge
\cite{yang,kobesmirl,beckerss,lss,jbauer} \textquotedblleft
solve\textquotedblright\ this problem by attaching a unitary operator to all
states, including non-interacting states: $\widetilde{\Phi}=U\Phi$,
$\widetilde{\Psi}=U\Psi$. All $U$ and $U^{-1}$ operators exactly cancel in the
matrix element, and the transition amplitude is unchanged. This is what leads
to the property \textquotedblleft gauge-invariant formalism\textquotedblright%
\ sometimes ascribed to the length gauge. However, this procedure amounts to
an identity or to a change of \textit{quantum picture}, but not to a gauge transformation.

\section{Fundamental contrasts in the applicability of fields and potentials}

The first example to be presented is the very basic one of the impossibility
of expressing the Schr\"{o}dinger equation directly in terms of electric and
magnetic fields, which should be possible if fields are truly more fundamental
than potentials. Other direct examples of difficulties posed by the assumption
of the fundamental importance of fields are shown, many of them long employed
unnoticed within the strong-field community.

\subsection{Schr\"{o}dinger equation}

The Schr\"{o}dinger equation%
\begin{equation}
i\hslash\partial_{t}\Psi\left(  t\right)  =H\Psi\left(  t\right)  , \label{ak}%
\end{equation}
when viewed as a statement in a Hilbert space (that is, without selecting a
representation such as the configuration representation or the momentum
representation) states that the effect of rotating a state vector $\Psi\left(
t\right)  $ by the operator $H$ within the Hilbert space produces the same
effect as differentiating the vector with respect to time (multiplied by
$i\hslash$). Time $t$ is an external parameter upon which the state vectors
depend, which accounts for why Eq. (\ref{ak}) specifies $t$ as a label
independent of the Hilbert space. A unitary transformation preserves this
equivalence. This can be stated as%
\begin{equation}
i\hslash\partial_{t}-\widetilde{H}=U\left(  i\hslash\partial_{t}-H\right)
U^{-1}. \label{al}%
\end{equation}
Since Eq. (\ref{al}) can be written as%
\begin{equation}
\widetilde{H}=UHU^{-1}+U\left(  i\hslash\partial_{t}U^{-1}\right)  ,
\label{am}%
\end{equation}
this shows explicitly that the Hamiltonian operator does not transform
unitarily if there is any time dependence in $U$.

An important gauge transformation is that introduced by G\"{o}ppert-Mayer
\cite{gm}, widely employed in the AMO community. This transformation is given
by%
\begin{equation}
U_{GM}=\exp\left(  \frac{ie}{\hslash c}\mathbf{r\cdot A}\left(  t\right)
\right)  , \label{an}%
\end{equation}
which depends explicitly on time when $\mathbf{A}\left(  t\right)  $ describes
a laser field within the dipole approximation, meaning that Eq. (\ref{am}) is consequential.

The fact that, in general,%
\begin{equation}
\widetilde{H}\neq UHU^{-1}, \label{ao}%
\end{equation}
is the explanation for the curious result to be found in many papers (for
example, Refs. \cite{yang,kobesmirl,beckerss,lss,jbauer}) that the
$\mathbf{r\cdot E}$ potential is a preferred potential. If any other potential
is employed in solving the Schr\"{o}dinger equation, then the claim is made
that a transformation factor must be employed even on a non-interacting state.
There is a logical contradiction inherent in the requirement that a
non-interacting state must incorporate a factor that depends on an
interaction, but the list of published papers that accept this premise is much
longer than the salient examples cited here. The underlying problem is the
assumption that a gauge transformation transforms the Hamiltonian unitarily.
That problem exists in all of the references just cited, although it is
usually submerged in complicated manipulations. It is especially clear in Ref.
\cite{jbauer}, where it is specified that all operators $O$ transform under a
gauge transformation according to the unitary-transformation rule%
\begin{equation}
\widetilde{O}=UOU^{-1},\qquad U^{-1}=U^{\dag}. \label{ap}%
\end{equation}
That specification is applied to $H$, in violation of the condition
(\ref{am}), and to the interaction Hamiltonian $H_{I}$, with no explanation
for how it is possible to gauge-transform from the length-gauge interaction to
any other gauge in view of the absence of operators in the scalar potential
$\mathbf{r\cdot E}$. In the scheme proposed in Refs.
\cite{yang,kobesmirl,beckerss,lss,jbauer}, if the problem is initially
formulated in the context of the $\mathbf{r\cdot E}$ potential, it is never
possible to transform to any other gauge. This explains the use of the phrase
\textquotedblleft gauge-invariant formulation\textquotedblright\ with respect
to $\mathbf{r\cdot E}$ to be found in some published works.

\subsection{Locality and nonlocality}

Fields are derived from potentials by the calculus process of differentiation,
as exhibited in Eq. (\ref{a}). Differentiation is carried out at a point in
spacetime. It is \textit{local}. If potentials are to be expressed from
fields, that requires integration, consisting of information from a range of
spacetime values; it is \textit{nonlocal}. The fact that the Schr\"{o}dinger
equation requires the local information from potentials, and cannot be
described by fields without inferring nonlocality in spacetime, is direct
evidence that potentials are more fundamental than fields.

\subsection{Ambiguity in the electromagnetic field tensors}

The basic field tensor of electrodynamics is defined as%
\begin{equation}
F^{\mu\nu}=\partial^{\mu}A^{\nu}-\partial^{\nu}A^{\mu}. \label{aj1}%
\end{equation}
It is important to note that this expression is in terms of the derivatives of
potentials rather than the potentials themselves. Thus it is not surprising
that the Lorentz invariant found from the inner product of $F^{\mu\nu}$ with
itself yields an expression in terms of fields:%
\begin{equation}
F^{\mu\nu}F_{\mu\nu}=2\left(  \mathbf{B}^{2}-\mathbf{E}^{2}\right)  .
\label{aj2}%
\end{equation}
A dual tensor can be defined as%
\begin{equation}
G^{\mu\nu}=\frac{1}{2}\epsilon^{\mu\nu\rho\lambda}F_{\rho\lambda}, \label{aj3}%
\end{equation}
where $\epsilon^{\mu\nu\rho\lambda}$ is the completely asymmetric fourth-rank
tensor. (The conventions of Jackson \cite{jackson} are being employed.) The
inner product of the basic and dual tensors gives a second Lorentz invariant:%
\begin{equation}
G^{\mu\nu}F_{\mu\nu}=-4\mathbf{B\cdot E}, \label{aj4}%
\end{equation}
also in terms of fields. The two Lorentz invariants%
\begin{equation}
\mathbf{E}^{2}-\mathbf{B}^{2},\quad\mathbf{E\cdot B} \label{aj5}%
\end{equation}
are said to characterize the electrodynamic environment.

An important special case is that of transverse, propagating fields, where%
\begin{equation}
\mathbf{E}^{2}-\mathbf{B}^{2}=0,\quad\mathbf{E\cdot B}=0. \label{aj6}%
\end{equation}
The properties (\ref{aj6}) lead to radiation fields as sometimes being called
\textquotedblleft null fields\textquotedblright. (The terms radiation field,
propagating field, transverse field, plane-wave field, are here used
interchangeably.) Radiation fields propagate at the speed of light in vacuum,
and they have the unique character that, after initial formation, they
propagate indefinitely in vacuum without the presence of sources.

However, the invariants (\ref{aj6}) are not unique to radiation fields; they
apply also to \textquotedblleft constant crossed fields\textquotedblright.
That is, it is always possible to generate static electric and magnetic fields
of equal magnitude that are perpendicular to each other, and will thus possess
zero values for both of the Lorentz invariants of the electromagnetic field.
Constant crossed fields do not propagate, and they cannot exist without the
presence of sources. They are unrelated to radiation fields despite sharing
the same values of the Lorentz invariants. Most importantly, constant crossed
fields cannot be considered as the zero-frequency limit of radiation fields,
as they are sometimes described \cite{nikrit66,ritus85}. All radiation fields
propagate at the speed of light for all frequencies, no matter how low. There
is no possible zero-frequency static limit \cite{hrtun}.

There is no ambiguity when radiation fields and constant crossed fields are
expressed in terms of their potentials. Radiation-field potentials possess the
periodicity inherent in trigonometric dependence on the $\varphi$ of Eq.
(\ref{ab}). This is unrelated to the $\phi=-\mathbf{r\cdot E}_{0}$ potential
of (\ref{o}) for a constant electric field $\mathbf{E}_{0}$, and
$\mathbf{A}=-\left(  \mathbf{r\times B}_{0}\right)  /2$ for a constant
magnetic field $\mathbf{B}_{0}$, both of which require source terms for their existence.

\subsection{Lorentz force}

The force exerted on a particle of charge $q$ moving with velocity
$\mathbf{v}$ in a field with electric and magnetic components $\mathbf{E}$ and
$\mathbf{B}$ is given by the Lorentz force expression%
\begin{equation}
\mathbf{F}=q\left(  \mathbf{E}+\frac{\mathbf{v}}{c}\times\mathbf{B}\right)  .
\label{aj7}%
\end{equation}
In a plane-wave field, the electric and magnetic fields are of equal
magnitude: $\left\vert \mathbf{E}\right\vert =\left\vert \mathbf{B}\right\vert
$. Thus, under conditions where $\left\vert \mathbf{v}\right\vert /c\ll1$, the
magnetic component of the force is minor as compared to the electric
component. The implication is that, as the field frequency declines, the
motion-related magnetic component reduces to an adiabatic limit. This concept
of adiabaticity justifies the complete neglect of the magnetic field that is a
key element of the dipole approximation, applied within the AMO community in
the form:%
\begin{equation}
\mathbf{E}=\mathbf{E}\left(  t\right)  ,\quad\mathbf{B}=0. \label{aj8}%
\end{equation}
The adiabaticity line of reasoning suggests a so-called adiabatic limit where
the field frequency declines to zero, and the plane wave field behaves as a
constant crossed field that satisfies the plane-wave condition (\ref{aj6}).

The entire line of reasoning that involves the concepts of adiabaticity,
adiabatic limit, and a zero-frequency limit for plane-wave fields is
field-based and erroneous \cite{hr101,hrtun}. When the problem is treated in
terms of potentials, it becomes clear that $v/c$ approaches unity for very
strong fields even when the frequency is very low, and the magnetic force
becomes equivalent to the electric force for very strong fields.

\subsection{Critical field}

The \textquotedblleft critical field\textquotedblright\ in electrodynamics is
related to the spontaneous breakdown of the vacuum into electron-positron
pairs. The critical field is defined as the electric field strength at which
the $\pm mc^{2}$ limits for the rest energies of electron and positron in a
particle-hole picture are \textquotedblleft tilted\textquotedblright\ by the
electric field to allow tunneling between positive and negative energy states
when the spatial limits of the tunnel are separated by an electron Compton
wavelength. This type of pair production is called Sauter-Schwinger pair
production \cite{sauter,schwinger}. It is fundamentally different from
Breit-Wheeler pair production \cite{bw,hrdiss,hr62} to be discussed below.

The reason for the fundamental difference is that Sauter-Schwinger pair
production is a phenomenon due to electric fields and Breit-Wheeler is due to
plane-wave fields. The distinction between pure electric fields and plane-wave
fields could hardly be more clear, since both types of fields have unique
gauge choices that are contrasting.

The critical field is often mentioned as a goal of strong-field laser
facilities, which is basically a \textit{non-sequitur}. The critical field
applies only to electric fields and lasers produce plane-wave fields. Even if
a laser field were sufficiently intense that its electric component had the
magnitude of the QSE critical field, pair production from the vacuum cannot
occur because the photons of the laser field convey momentum. A
counter-propagating plane-wave field is necessary to satisfy momentum
conservation as well as energy conservation in the production of pairs from
the vacuum. This is then the two-fields Breit-Wheeler process, which is
unrelated to the single-field Sauter-Schwinger process.

\subsection{Pair production from the vacuum}

For many years the stated ultimate goal of large-laser programs was to achieve
a laser intensity such that the electric component of the laser is equal to
the critical field discussed in the preceding subsection This magnitude of
electric field corresponds to an intensity of about $4.6\times10^{29}W/cm^{2}$.

The problem is that the Schwinger limit applies \textit{only} to electric
fields. It does not apply to laser fields \cite{hrspie}, as explained above.
The only way to produce momentum balance with a laser field while still
producing pairs from the vacuum is to have the laser beam collide with
oppositely-directed photons, as proposed in Ref. \cite{hr71}. This was
predicted to be done on a practical basis at a linear accelerator facility
such as that at SLAC\ (Stanford Linear Accelerator Center), with a laser
intensity of only slightly greater than $10^{18}W/cm^{2}$. An important note
is that the prediction of Ref. \cite{hr71} was for the use of the
then-important ruby laser at a different pulse length and a different energy
of the energetic electron beam used to produce the countervailing photon
field. The predicted threshold of 25 photons from the laser field in Ref.
\cite{hr71} is altered to 5 photons for the parameters of the experiment that
was actually done at SLAC in 1997 \cite{burke}. The theoretical prediction of
an effective threshold intensity of about $10^{18}W/cm^{2}$ is maintained
because of the essential independence of the laser frequency that was remarked
in Ref. \cite{hr71}.

(One \textit{caveat} about the experiment is that it was reported as a
high-order perturbative result, whereas it is readily shown to be at an
intensity beyond the radius of convergence of perturbation theory \cite{epjd}.
A second problem is that it was described as \textquotedblleft light-by-light
scattering\textquotedblright, which is a different process altogether. Feynman
diagrams of these processes have electron and positron as emergent particles
in a pair production process, while light-by-light scattering has emergent photons.)

The striking difference between the $4.6\times10^{29}W/cm^{2}$ required to
attain a laser electric component equal to the critical field and the actual
$1.3\times10^{18}W/cm^{2}$ required for the SLAC experiment is evidence of a
misplaced focus on the electric field required for vacuum pair production. The
required intensity of the laser field depends on the properties of the
counter-propagating field, but it is never as large as the Sauter-Schwinger
critical field.

\subsection{Low frequency limit of a plane-wave field}

It is conventional to view low-frequency laser-induced phenomena from the
standpoint of the Lorentz force expression of Eq. (\ref{aj7}). In a plane-wave
field, the electric and magnetic fields are of equal magnitude: $\left\vert
\mathbf{E}\right\vert =\left\vert \mathbf{B}\right\vert $. Thus, under
conditions where $\left\vert \mathbf{v}\right\vert /c\ll1$, the magnetic
component of the force is minor as compared to the electric component. The
implication is that, as the field frequency declines, the motion-related
magnetic component reduces to an adiabatic limit. This concept of adiabaticity
appears to justify the complete neglect of the magnetic field that is a key
element of the dipole approximation, applied within the AMO community in the
form given in Eq. (\ref{aj8}). Adiabaticity suggests a so-called adiabatic
limit, where the field frequency declines to zero, and the plane wave field
behaves as a constant crossed field that satisfies the plane-wave condition
(\ref{aj6}).

The entire line of reasoning that involves the concepts of adiabaticity,
adiabatic limit, and a zero-frequency limit for plane-wave fields is
field-based and erroneous \cite{hr101,hrtun}. When the problem is treated in
terms of potentials, it becomes clear that $v/c$ approaches unity for very
strong fields. The magnetic force becomes equivalent to the electric force for
very strong fields.

Analysis from a potentials standpoint is in stark contrast to a fields-based approach.

The ponderomotive potential energy of a charged particle in a plane-wave field
is a fundamental property of the particle \cite{hrup}, and it becomes
divergent as the field frequency approaches zero. The immediate consequence is
that the limit $\omega\rightarrow0$ causes the dipole approximation to fail
\cite{hr101,hrtun}, and corresponds to an extremely relativistic environment
\cite{hrup}. This contradicts maximally the field-based conclusions.

The ponderomotive energy $U_{p}$ is given by%
\begin{equation}
U_{p}=\frac{q^{2}}{2mc^{2}}\left\langle \left\vert A^{\mu}A_{\mu}\right\vert
\right\rangle , \label{aj9}%
\end{equation}
where the angle brackets denote a cycle average, and the absolute value needs
to be indicated because the 4-vector potential $A^{u}$ is a spacelike 4-vector
and its square is thus negative with the metric being employed. The
ponderomotive energy is based on potentials. When expressed in terms of field
intensity $I$, $U_{p}$ behaves as%
\begin{equation}
U_{p}\sim I/\omega^{2}, \label{aj10}%
\end{equation}
which explains the relativistic property of the charged particle as the
frequency approaches zero.

\subsection{Extremely-low-frequency radio waves}

A central issue in this article is that transverse fields and longitudinal
fields are fundamentally different, even when they seem to have some
properties in common. An effective example is the matter of communicating with
submerged submarines with extremely low frequency (ELF) radio waves. The U. S.
Navy operated such a system \cite{wikisanguine}, designed to communicate with
submarines submerged at depths of the order of $100m$ over an operational
range of about half of the Earth's surface. The point of using extremely low
frequencies is the large skin depth in a conducting medium (seawater) that can
be achieved. What is most remarkable about the system is that the frequency of
$76Hz$ that was used has a wavelength of about $4\times10^{6}m$, which is
approximately $0.62$ times the radius of the Earth. The system could convey an
intelligible signal over about half of the Earth's surface. Considering the
length of the submarine (about $100m$) to be the size of the receiving
antenna, this means that the wavelength to antenna-length ratio is about
$4\times10^{4}$. That is, the received radio wave is constant over the entire
length of the receiving antenna to a very high degree of accuracy. A
nearly-constant electric field cannot be detected at a distance of half of our
planet away from its source; a nearly-constant electric field cannot penetrate
$100m$ through a conducting medium; and so a nearly-constant electric field
cannot convey an intelligible signal in the presence of all of these extremely
effective barriers. A longitudinal field is fundamentally different from a
transverse field.

The proposed \textquotedblleft local constant-field
approximation\textquotedblright\ (LCFA) for relativistic laser effects
\cite{ritus85,heidrmp} is based on the presumed similarity of low-frequency
laser fields to constant crossed fields. This was shown earlier in this
manuscript to be a meaningless association. The application involved in the
communication with submerged submarines is a practical demonstration that the
LCFA is not a valid concept.

\section{Hierarchy of physical effects}

In basic calculus, if a function $f\left(  x\right)  $ is known, then so is
$df/dx$; differentiation is local. If the derivative is all that is known, the
process of integration to find $f\left(  x\right)  $ requires the knowledge of
$df/dx$ over a range of values; integration is nonlocal. In electromagnetism,
potentials are the analog of $f\left(  x\right)  $ and fields correspond to
$df/dx$. The example of the Schr\"{o}dinger equation shows that knowledge of
potentials is primary and fields are secondary. This ordering can be extended
to other physical quantities. For example, the Lorentz force expression in Eq.
(\ref{aj7}) relates forces directly to fields, meaning that forces are
secondary. A further implication is that Newtonian mechanics, expressed in
terms of forces, is secondary to Lagrangian and Hamiltonian mechanics that are
expressed in terms of energies; that is, in terms of potentials. It is no
accident that mechanics textbooks show that formalisms based on system
functions like the Lagrangian or Hamiltonian infer the Newtonian formalism,
but to go in the reverse direction requires an extrapolation of concepts.

This ordering, or hierarchy, of physical quantities has consequences in
problems that go beyond simple electric-only or magnetic-only fields. In
laser-related problems where both electric and magnetic fields are involved,
it is possible to arrive at invalid inferences if only secondary influences
are regarded as controlling. An example is the common practice of viewing
electric fields as being the dominant quantities in long-wavelength
circumstances where the dipole approximation appears to be valid. This
disguises the fact that extremely low frequencies can lead to relativistic
behavior, where electric fields supply inadequate information
\cite{hr101,hrtun}, and inappropriate concepts such as adiabaticity exist.

A cogent example is the critical field. A widely used \textquotedblleft
strong-field QED parameter\textquotedblright\ is simply the ratio of the
electric field to the critical field. This has meaning only for
static-electric or for QSE fields. For strong-field laser applications,
relevant intensity parameters are all ratios of energies. See, for example,
the section entitled \textquotedblleft Measures of intensity\textquotedblright%
\ in Ref. \cite{hrrev}. (The early, but still useful review article by Eberly
\cite{eberlyreview} is also instructive in this regard.) In particular, the
\textquotedblleft free-electron intensity parameter\textquotedblright%
\ $z_{f}=2U_{p}/mc^{2}$ is known to measure the coupling between an electron
and a nonperturbatively intense plane-wave field \cite{hrdiss,hr62,hr62b,hrup}%
, replacing the fine-structure constant of perturbative electrodynamics.

A special insight arises when the $z_{f}$ parameter is related to the
fine-structure constant $\alpha$ \cite{hrup}:%
\begin{equation}
z_{f}=\alpha\rho\left(  2\lambda\lambdabar_{C}^{2}\right)  , \label{aj11}%
\end{equation}
where $\rho$ is the number of photons per unit volume, and $2\lambda
\lambdabar_{C}^{2}$ is approximately the volume of a cylinder of radius equal
to the electron Compton wavelength $\lambdabar_{C}$ and a length given by the
wavelength $\lambda$ of the laser field. That is, it appears that all of the
photons within the cylinder participate in the coupling between the electron
and the laser field. The electron Compton wavelength is the usual interaction
distance for a free electron, but the wavelength can be a macroscopic
quantity. The \textquotedblleft strong-field physics\textquotedblright\ that
arises from the application of the Volkov solution to problems involving the
interaction of very intense radiation fields with matter
\cite{sengupta,hrdiss,hr62}, thus appears to be the bridge connecting quantum
electrodynamics with the classical electrodynamics of Maxwell.

Other examples of the hazards of excessive dependence on secondary quantities
have been mentioned above in the context of the assumed dominance of the
electric component of the Lorentz force at low frequencies, and the inference
from the Lorentz invariants that constant crossed fields are related to
propagating fields.

\section{Failure of perturbation theory}

Perturbation theory has been the cornerstone of QED since the Nobel-prize
winning work of Feynman, Schwinger, and Tomonaga. The radius of convergence of
perturbation theory was examined in depth a long time ago \cite{hrdiss}. The
motivation for the study was the demonstration by Dyson \cite{dyson} that,
despite its remarkable quantitative success, standard covariant QED has a zero
radius of convergence for a perturbation expansion. The question about whether
a strong-field theory based on the Volkov solution of the Dirac equation
\cite{volkov} could be convergent was the motivation for Ref. \cite{hrdiss}.
The answer was affirmative, but with intensity-dependent singularities in the
complex coupling-constant plane that limited the convergence. (This explains
the $z_{f}$ terminology, since $z_{f}$ was allowed to be complex, and the
quantity $z$ is often used for complex numbers.) Physically, convergence fails
whenever the field intensity is high enough to cause a channel closing. That
is, if a process requires a certain threshold energy to proceed in a
field-free process as measured by some photon number $n_{0}$, that threshold
energy will be increased due to the need for a free electron to possess the
ponderomotive energy $U_{p}$ in the field. When $U_{p}$ is large enough to
cause $n_{0}$ to index up to $n_{0}+1$, that is called a channel-closing, and
it marks a sufficient condition for the failure of perturbation theory. The
upper limit on perturbation theory is therefore set by%
\begin{equation}
U_{p}<\hslash\omega,\text{ \ or \ }I<4\omega^{3}, \label{aj12}%
\end{equation}
where the last expression is in atomic units, using the equivalence
$U_{p}=I/4\omega^{2}$ in atomic units, and $I$ is the laser intensity.

Although the original convergence investigation was done for free electrons,
the same limit was found for atomic ionization \cite{hr80}.

The relevance of using an index based on the ratio of the electric field to
the critical field as an \textit{ad hoc }limit on perturbation theory for
laser effects is strongly questioned here. Such a basic matter as the
applicability of perturbation methods to the treatment of the effects of
radiation fields is governed by primary quantities like the energies $U_{p}$
and $mc^{2}$, and not by secondary quantities like the magnitude of the
electric field.

\section{Overview}

The Aharonov-Bohm effect introduced a major change in electrodynamics because
it showed that potentials are indeed more fundamental than fields. However,
although the effect relates to a quantum phenomenon, it has had little effect
on the way in which quantum mechanics is employed. The ascendancy of
potentials over fields as described above is much more consequential,
especially for strong-field phenomena. A simple summarizing statement is that
electromagnetic scalar and vector potentials convey more physical information
than the electric and magnetic fields derivable from them.

Of special importance is the fact that two cases have been identified where
there can be only one acceptable set of potentials, and these two cases are of
very wide practical scope. One case is the constant electric field, which is
exactly or approximately applicable to a wide variety of AMO and
condensed-matter applications. The unique acceptable potential for static
electric fields is the length-gauge, or $\mathbf{r\cdot E}$ potential, and
this is already widely employed for constant or slowly varying electric-field applications.

The other case with a unique allowable gauge is the propagating-field case,
which is of profound importance in laser-based experiments. Such experiments
constitute an increasingly large proportion of AMO activities. An essential
reminder is that only propagating fields can persist in the absence of
sources, so that virtually all laser-based experiments occur in a
superposition of propagating-field components. The radiation gauge (or Coulomb
gauge) is the only gauge employable without risking the creation of hidden
errors due to improper gauge choice. It is unfortunate that, in strong-field
laser applications, the widespread use of the dipole approximation introduces
precisely that hazard of hidden errors that affects both practical
calculations and qualitative insights into the behavior of physical systems
subjected to laser fields.

It has been shown that gauge transformations are not, in general, unitary.
This has never previously been reported, and it can lead to further errors in
addition to the important example explored in Section III above about the
putative universality of the length gauge.

The concept of primary and secondary physical quantities has been introduced,
with over-dependence on a secondary quantity like the electric field having
the capability of leading to important misconceptions.

The criterion of a critical field for longitudinal fields has no relevance to
the transverse field of laser-induced processes.

The appearance of mixed quantum and classical quantities in ascertaining the
limits on perturbative methods in the application of strong-field theories
identifies strong-field physics as the bridge connecting quantum
electrodynamics and the classical electromagnetism of Maxwell.

\end{document}